\documentclass{amsart}
\usepackage{amssymb}
\usepackage{graphicx}
\usepackage{url}
 \newtheorem{theorem}{Theorem}
 \newtheorem{lemma}[theorem]{Lemma}
 \newtheorem{proposition}[theorem]{Proposition}
 \newtheorem{corollary}[theorem]{Corollary}
 \newtheorem{remark}[theorem]{Remark}
 \newtheorem{example}[theorem]{Example}
 \newtheorem{definition}[theorem]{Definition}
 \newtheorem{conjecture}[theorem]{Conjecture}

 \newtheorem{question}[theorem]{Question}

\newcommand{\bpr}{\begin{proof}}
\newcommand{\epr}{\end{proof}}
\newcommand{\beq}{\begin{equation}}
\newcommand{\eeq}{\end{equation}}
\newcommand{\bThm}{\begin{theorem}}
\newcommand{\eThm}{\end{theorem}}
\newcommand{\blem}{\begin{lemma}}
\newcommand{\elem}{\end{lemma}}
\newcommand{\bpro}{\begin{proposition}}
\newcommand{\epro}{\end{proposition}}
\newcommand{\bcor}{\begin{corollary}}
\newcommand{\ecor}{\end{corollary}}
\newcommand{\brem}{\begin{remark}}
\newcommand{\erem}{\end{remark}}
\newcommand{\bexa}{\begin{example}}
\newcommand{\eexa}{\end{example}}
\newcommand{\bdf}{\begin{definition}}
\newcommand{\edf}{\end{definition}}
\newcommand{\bcon}{\begin{conjecture}}
\newcommand{\econ}{\end{conjecture}}
\newcommand{\bque}{\begin{question}}
\newcommand{\eque}{\end{question}}

\newcommand{\comment}[1]{}

\title{Causality and quantum theory}
\author{Blake K. Winter}
\address{Oakland University, bkwinter@oakland.edu}


\date{}

\begin{document}
\thispagestyle{empty}

\begin{abstract}
We begin with a brief summary of issues encountered involving causality in quantum theory, placing careful emphasis on the assumptions involved in results such as the EPR paradox and Bell's inequality. We critique some solutions to the resulting paradox, including Rovelli's relational quantum mechanics and the many-worlds interpretation. We then discuss how a spacetime manifold could come about on the classical level out of a quantum system, by constructing a space with a topology out of the algebra of observables, and show that even with an hypothesis of superluminal causation enforcing consistent measurements of entangled states, a causal cone structure arises on the classical level. Finally, we discuss the possibility that causality as understood in classical relativistic physics may be an emergent symmetry which does not hold on the quantum level.
\end{abstract}

\maketitle
%
\section{Quantum theory}
%
Recall that a quantum system may be described by a Hilbert space $H$ (for technical reasons, this must often be a rigged Hilbert space, \cite{Antoine, Madrid, TT} together with some subalgebra $O$ of the self-adjoint operators on $H$. Operators in $O$ correspond to observables. For example, the spin of a spin-$1/2$ particle is described by a two-dimensional Hilbert space with an orthonormal basis of unit vectors given by (using Dirac's bra-ket notation) $\left| u \right\rangle$ and $\left| d \right\rangle$, the up and down states respectively. A \emph{state} of the system is a unit vector in $H$. According to the usual rules of quantum mechanics. When an observable $A\in O$ is measured, the possible outcomes are given by the spectrum of $A$. If $a_i$ is an eigenvalue for $A$, and the system is in the state $\left| \Psi \right\rangle$ then the probability of obtaining $a_i$ as the result of a measurement of $A$ is given by $|\left\langle \Psi \right|P^t_{a_i} P_{a_i}\left| \Psi \right\rangle|^2$, where $P_{a_i}$ is the projection operator onto the eigenspace of $A$ corresponding to the eigenvalue $a_i$. The quantity $\left\langle \Phi \right| \Psi \rangle$ is called the \emph{probability amplitude} of finding the system in the state $\left| \Phi \right\rangle$ when we measure it while it is in the state $\left| \Psi \right\rangle$.

According to the usual rules of quantum mechanics, a system normally evolves in a unitary fashion. That is, the state $\left| \Psi(t) \right\rangle$ is given by $U(t)\left| \Psi(0) \right\rangle$ for some unitary function $U(t)$. However, when a system is measured, the state changes in a non-unitary fashion: if a measurement of the observable $A$ while the system is in the state $\left| \Psi \right\rangle $yields the result $a_i$, then the system changes to the state $P_{a_i}\left| \Psi \right\rangle$. This is the usual \emph{axiom of measurement} in quantum theory. The difficulty in interpreting quantum mechanics is that the resulting probabilities are not additive, as we might expect in classical probability theory; rather, if a state $\left| \Psi \right\rangle=a\left| \Psi_1 \right\rangle + b\left| \Psi_2 \right\rangle$, then the probability of finding the system in the state $\left| \Phi \right\rangle$ will be given by $|\langle \Phi | ( a\left| \Psi_1 \right\rangle + b\left| \Psi_2\right\rangle)|^2$, that is, we are adding the probabilities amplitudes rather than the probabilities. This gives rise to the well-known phenomenon of quantum interference, as well as to the well-known difficulty in interpreting the concept of measurement in quantum mechanics.

\section{EPR, Bell, and causality}

The Einstein-Podolsky-Rosen paradox in quantum theory is a result which shows that ordinary quantum theory, assuming that our universe is a single universe, must either involve hidden variables, or else there must be a mechanism which involves superluminal causation. The original result is found in \cite{EPR}. Note that the assumption that our universe is a single universe simply means that all observers involved are single observers, and they are the same observers at each point in the history considered, and it does not say anything about whether parallel universes exist or not. This assumption is denied by the many-worlds interpretation, \cite{JB}, which we will discuss in Sec. \ref{MWsec}.

\begin{theorem}
Quantum theory in a single universe without hidden variables requires a superluminal mechanism to enforce consistent measurements, \cite{EPR}.
\end{theorem}.

\bpr As an example, consider two spin-$1/2$ particles. The Hilbert space describing their spins will be the tensor product of two copies of the Hilbert space with basis $\left| u \right\rangle$ and $\left| d \right\rangle$. Thus, the Hilbert space will have a basis consisting of four vectors: $\left| u_0 \right\rangle\left| u_1 \right\rangle$, $\left| u_0 \right\rangle\left| d_1 \right\rangle$, $\left| d_0 \right\rangle\left| u_1 \right\rangle$, $\left| d_0 \right\rangle\left| d_1 \right\rangle$. Suppose these particles are produced at a certain point in spacetime and sent off in opposite directions at close to the speed of light, hitting two detectors which are separated by a spacelike interval. Suppose furthermore that their spins, prior to hitting the detectors, are described by the vector $(\frac{1}{\sqrt{2}}(\left| u_0 \right\rangle\left| u_1 \right\rangle + \left| d_0 \right\rangle\left| d_1 \right\rangle)$. Then when results are compared, either both particles are found to have been spin up, or both have been found to be spin down. But this requires either that there be local hidden variables telling the particles which state they should collapse into upon measurement, or else some superluminal mechanism to enforce their correlation. \epr

On the other hand, Bell showed that local hidden variables are incapable of reproducing the results of quantum mechanics.

\begin{theorem}
No local hidden variable theory can reproduce the measured results of quantum mechanics.
\end{theorem}

See \cite{Bell} for a proof of this theorem, and \cite{Belltest} for an experiment demonstrating that the quantum mechanical results are experimentally verified.

Bell's theorem seems to frequently be cited as proof that hidden variables cannot be local (though non-local hidden variables, or hidden variables with superluminal causal enforcement mechanisms, can reproduce the results of quantum mechanics, e.g. \cite{DB}). However, it is often overlooked that the EPR result says that without hidden variables, there must still be some nonlocal enforcement mechanism, provided that our universe remains a single universe through history, and provided that correlations require some real causal mechanism of enforcement.

\section{Relational Quantum Mechanics and Many-Worlds}\label{MWsec}

Here we will discuss two methods for trying to resolve the above issue. The first, relational quantum mechanics, attempts to do this in a single universe. It thus ends up violating the supposition that every correlation has some real mechanism of enforcement underlying it, and appears logically unsound. The second does away with the assumption that our universe is a single universe.

We begin with the notion of relational quantum mechanics, which relies on the idea that the split of the physical universe into quantum systems and classical observers is relative or \emph{relational} and has been suggested by Rovelli, \cite{CR}. According to this interpretation, a state vector exists only relative to an observer. Observer A may be a part of observer B's state vector (and vice verse). There are two problems with this approach. First, it does not define what systems will actually give rise to observers. Therefore, it depends, like the Copenhagen interpretation, on arbitrarily inserting observers into the physical universe. However, it is possible that this problem could be eventually resolved in a satisfactory manner. Second, it does not explain why two observers will agree whenever they compare experimental results. As an example, suppose there are a pair of spin-$1/2$ particles, a and b, which are entangled so they are both spin up or both spin down. Suppose observer A measures the spin of particle a and observer B measures the spin of particle B. Then according to observer A, measuring particle a causes the state vector to collapse to a state where particle b has exactly one spin, while observer B sees measuring particle b to cause his state vector to collapse particle a to a state with only one spin. When A and B compare their results, A sees B's measurement as being caused by A's state vector, while B sees A's measurement as being caused by the collapse of B's state vector. Rovelli argues that this is no different than the case of chronological ordering in special relativity, in which two observers may disagree about which event precedes another. Therefore, Rovelli is comfortable with the fact that A and B disagree about the cause of correlation between A's results and B's results.

However, we maintain that the two cases are quite different. In special relativity, chronological order is no longer fixed, but there is still a well-defined \emph{causal} order between events. Two observers in special relativity will agree about the causes of a chain of events, even though they might disagree on the chronological order. According to our understanding of Rovelli's relational quantum mechanics, however, the question "what \emph{actually} causes there to be a correlation between A's measurements and B's measurements" has no answer. A and B will disagree about the cause. \emph{Relational quantum mechanics is therefore incompatible with the philosophical tenet that there is an objective or real cause for the correlation between the measurements of the two observers.}

The only method that we can see to try to salvage both the relational interpretation and the objectivity of causes is to add in a many-worlds type interpretation, \cite{JB}. In order to avoid any kind of spacelike causes, we might posit that when observer A goes to meet observer B, the act of comparing results causes observer A to branch into a universe with a compatible version of observer B. Note that we cannot say that A was already in that universe from the moment they made the first measurement, without returning to the need to enforce spacelike correlations. If A went into a universe where B already had a compatible measurement, then there is a spacelike correlation being enforced: which universe A goes into at the moment of initial measurement is determined by the measurement of (the many copies of) B at a spacelike interval away. Thus, we should instead hypothesize that A moves into the appropriate universe, with its compatible version of observer B, only when A and B meet to compare their results.

One problem with the many-worlds interpretation is quantum erasure. Quantum erasure occurs when a measurement is erased, which also results in the state vector no longer being collapsed. It has been experimentally observed, \cite{SW, KY}. However, according to the many-worlds interpretation, the new branch of the universe should only have the collapsed state vector. The fact of quantum erasure suggests that the old state vector must also somehow be available to the new branch of the universe, in case such an erasure happens.

Alternately, one might argue that real observations can never be erased, and that in the experiments done (such as in \cite{SW, KY}) there was no actual observation that was erased. This would necessitate the hypothesis that we might experimentally differentiate between real observations and mere entanglements between an experiment and the observational apparatus. As such, this would make predictions that are empirically different from those of traditional quantum mechanics.

It should be noted, of course, that the issue of quantum erasure is not a problem only for the many-worlds interpretation, but for any interpretation of quantum mechanics.

\section{Relativistic invariance on the classical level as an emergent property}

\subsection{Spacetime from quantum theory}\label{stqt}

The problem of quantizing general relativity is one which has had a rocky history, due to numerous technical problems. For some discussion of the attempts made so far, we refer the reader to \cite{TT}. Here we will merely note how a spacetime with a Lorentzian structure of causal cones could come about from a quantum theory.

Ordinarily, of course, a quantum theory is constructed on a spacetime manifold. However, we will here show how a quantum theory naturally gives rise to a space with a topology, and why this will, on the classical level, have causal cones built into the Poisson bracket structure.

Let us assume that we have some quantum theory with a (rigged) Hilbert space $V$ and an algebra of observables $O$, in the Heisenberg picture. Let $C$ be a the sub-algebra of $O$ which is physically interpreted as the Heisenberg-picture operators measuring field strengths, that is, "position" operators (as opposed to, for example, momentum operators). We will create a topological spacetime from this in the following manner. Let $M$ be a set whose elements consist of minimal non-empty intersections of complete sets of commuting observables in $C$ (these complete sets of commuting observables are essentially Cauchy data). Note that this associates elements of $C$ with elements of $M$: each element of $C$ will belong to a set which is an element of $M$. The elements of $M$ will form the points of the spacetime. To topologize $M$, we note that a complete set of commuting observables should define a (spacelike) hypersurface. Also, given a point $x\in M$, the set of points $y\in M$ such that the elements of $C$ associated with $y$ commute with the elements of $C$ associated with $x$ should form an open set. Finally, each point should be a closed set. We may give $M$ the coarsest topology induced by these requirements.

\begin{remark}
If we apply this technique to non-relativistic quantum theories, we end up with $M$ being one-dimensional, and spacelike hypersurfaces consist of single points.
\end{remark}

In other words, if the intersection of any two distinct complete sets of commuting observables in $C$ is empty, then the result would be a one-dimensional topology. This suggests that even if there were a causal foliation of spacetime, nonetheless as long as space is not 0-dimensional, there would still be a Lorentzian-type of structure on the commutation relations of operators at various points. That is, an operator in $C$ associated with a given point $x$ will commute with operators on many different "spacelike" hypersurfaces in $M$. The points with which it does not commute would form an analog of a light cone. This would lead to a light cone type of structure on the classical level, where the commutators become Poisson brackets. Since measurements cannot themselves transmit classical information, it is not surprising that this would mean that on the classical level, the transmission of information would be restricted to these cones (barring some discovery of how to exploit the finer causal ordering of $M$ to send information faster).

\begin{remark}
This cone structure may have odd consequences in quantum gravity, since such a structure of causal cones will generally determine a metric up to a conformal factor, \cite{CC}. It is not clear how this may affect such theories; more work on this might yield interesting results.
\end{remark}

\subsection{Emergent symmetries}

It should be noted that a non-local enforcement mechanism need not involve an actual foliation of spacetime by spacelike slices to define a causal order. Indeed, one could try to posit that the enforcement mechanism were still in some sense relativistically invariant, as is done with the transactional interpretation of Cramer, \cite{JC}. However, there would still be some causal ordering beyond that given by the non-vanishing Poisson brackets of classical variables, for the following reason:

Suppose that, in the EPR setup, instead of simply having two particles, we used three, all entangled so they must all be found to be spin up or all be found to be spin down. Now suppose the third particle is sent in the same direction as the second particle, but slightly slower, so that it is measured at a point which is in the timelike future of the measurement of the second particle, but still spacelike separated from the measurement of the first particle. The measurement of the second particle causally precedes the measurement of the third particle, but involves a causal enforcement on the first particle's measurement. Therefore, the first particle's measurement appears to have to causally precede the measurement of the third particle, even though these are spacelike related events. This may not fully give a causal-ordering foliation of the entire spacetime manifold, but it does give rise to an ordering of some discrete subset of measurements, which is stronger than the classical causal ordering of those events.

\begin{theorem}
A superluminal enforcement mechanism for keeping measurements consistent, which behaves as indicated above, will result in a stronger causal ordering of measurements than would be expected classically.
\end{theorem}

Note: by \emph{stronger}, we mean here that two events which take place at points in spacetime which are not ordered by the classical partial ordering, will be ordered by this quantum causal ordering. Thus, the quantum causal ordering is a partial order relation which contains the classical partial ordering as a subset.

We cannot, however, rule out the possibility that one might do away with some of the assumptions above about how such a superluminal enforcement mechanism would behave, which might avoid this result. However, the assumptions made above seem reasonable.

However, we argue that this result is not necessarily surprising or problematic. Indeed, there are many instances where a symmetry on one scale is broken at a smaller scale, or where a classical symmetry is broken on the quantum scale.

For an example of the former, which does not even involve quantum physics, consider a physical model of billiard balls. Such a physical model can be made assuming that each ball is a perfect sphere, with the inherent rotational symmetry which that entails. However, on a microscopic level, the balls exhibit some degree of unevenness. This unevenness breaks the symmetry, but it is irrelevant at the scales involved in the model.

When it comes to quantum mechanics, it is not uncommon to find \emph{anomalies}. These are symmetries on the classical level which do not apply on the quantum level. For example, the \emph{chiral anomaly} appears in certain theories of fermions with chiral symmetry on the classical level: this symmetry may be broken on the quantum level. See \cite{Wein} for a discussion of this phenomenon.

Perhaps the most obvious example of a broken symmetry is the commutativity of position and momentum, which exists on the classical level, but not on the quantum level.

Therefore, it does not appear to us to pose a problem to hypothesize that there are superluminal causal enforcement mechanisms involved in measurements involving quantum entanglement, which do not necessarily obey classical relativistic symmetries. In particular, these violations of classical symmetries would only appear on the level of measurements, rather than on the level of the unitary evolution of the system. Therefore they would be expected to vanish classically, since in the classical limit, measurements become moot due to the deterministic nature of the classical limit. As seen in Sec. \ref{stqt}, a relativistic style of causal cones is a natural consequence of a the commutator relations of a quantum field theory. Therefore: even if we posit superluminal causal correlations for measurements, on the classical level we might expect to see Poisson brackets which correspond to a system of causal cones. Since measurements of entangled states cannot themselves be used to transfer information, this naturally results in classical information being restricted to transmission within said cones.

On the other hand, it is important to emphasize that the results of EPR logically require local hidden variables, a branching multiverse, or a superluminal mechanism to enforce consistency. Bell's result shows that the first of these options is empirically false, which logically leaves the latter two options. This is a fact which should not be overlooked, paradoxical as it may seem.

\end{document}